\documentclass[12pt,preprint]{aastex}
\usepackage{natbib}

\slugcomment{Submitted to Astrophysical Journal Letters}

\shorttitle{PSR J1856+0245}
\shortauthors{Hessels et al.}

\newcommand{\palfapsr}{PSR J1856+0245}
\newcommand{\palfapsrshort}{J1856+0245}
\newcommand{\hesssrc}{HESS J1857+026}
\newcommand{\ascaxray}{AX J185651+0245}
\newcommand{\asca}{{\it ASCA}}
\newcommand{\swift}{{\it Swift}}

\begin{document}

\title{\palfapsr: Arecibo Discovery of a Young,
Energetic Pulsar Coincident with the TeV $\gamma$-ray 
Source HESS J1857+026}

\author{J.~W.~T. Hessels$^{1,*}$, D.~J. Nice$^{2}$, B.~M.
Gaensler$^{3}$, V.~M. Kaspi$^{4}$, D.~R. Lorimer$^{5}$, D.~J.
Champion$^{6}$, A.~G. Lyne$^{7}$, M. Kramer$^{7}$, J.~M.
Cordes$^{8}$, P.~C.~C. Freire$^{9}$, F. Camilo$^{10}$, S.~M.
Ransom$^{11}$, J.~S. Deneva$^{8}$, N.~D.~R. Bhat$^{12}$, I.
Cognard$^{13}$, F. Crawford$^{14}$, F.~A. Jenet$^{15}$, L.
Kasian$^{16}$, P. Lazarus$^{4}$, J. van Leeuwen$^{17}$, M.~A.
McLaughlin$^{5}$, I.~H. Stairs$^{16}$, B.~W. Stappers$^{7}$, and A.
Venkataraman$^{9}$}

\affil{$^{1}$Astronomical Institute ``Anton Pannekoek,'' Univ.
of Amsterdam, 1098 SJ Amsterdam, The Netherlands} 

\affil{$^{2}$Physics Dept., Bryn Mawr College, Bryn Mawr, PA
19010} 

\affil{$^{3}$Institute of Astronomy, School of Physics, Univ.~of
Sydney, NSW 2006, Australia} 

\affil{$^{4}$Dept.~of Physics, McGill Univ., Montreal, QC H3A
2T8, Canada} 

\affil{$^{5}$Dept.~of Physics, West Virginia Univ.,
Morgantown, WV 26506}

\affil{$^{6}$ATNF-CSIRO, Epping, NSW 1710, Australia} 

\affil{$^{7}$Univ. of Manchester, Jodrell Bank Centre for Astrophysics, 
Alan Turing Building, Manchester M13 9PL, UK} 

\affil{$^{8}$Astronomy Dept., Cornell Univ., Ithaca, NY 14853} 

\affil{$^{9}$NAIC, Arecibo Observatory, PR 00612} 		                                       

\affil{$^{10}$Columbia Astrophysics Laboratory, Columbia Univ., New York, NY 10027}

\affil{$^{11}$NRAO, Charlottesville, VA 22903}               

\affil{$^{12}$Centre for Astrophysics and Supercomputing, Swinburne
Univ.~of Technology, Hawthorn, Victoria 3122, Australia} 

\affil{$^{13}$LPCE / CNRS, F-45071 Orleans Cedex 2, France}

\affil{$^{14}$Department of Physics and Astronomy, Franklin and
Marshall College, Lancaster, PA 17604}

\affil{$^{15}$Center for Gravitational Wave Astronomy, Univ. of
Texas at Brownsville, TX 78520} 

\affil{$^{16}$Dept.~of Physics and Astronomy, Univ.~of British
Columbia, Vancouver, BC V6T 1Z1, Canada} 

\affil{$^{17}$Astronomy Dept., Univ.~of
California at Berkeley, Berkeley, CA 94720} 

\affil{$^\ast$E-mail: J.W.T.Hessels@uva.nl} 

\begin{abstract} 
We present the discovery of the Vela-like radio pulsar \palfapsrshort\
in the Arecibo PALFA survey.  \palfapsr\ has a spin period of 81\,ms,
a characteristic age of 21\,kyr, and a spin-down luminosity $\dot{E} =
4.6 \times 10^{36}$\,ergs s$^{-1}$.  It is positionally coincident
with the TeV $\gamma$-ray source \hesssrc, which has no other known
counterparts. Young, energetic pulsars create wind nebulae, and more
than a dozen pulsar wind nebulae have been associated with
very-high-energy (100\,GeV$-$100\,TeV) $\gamma$-ray sources discovered
with the HESS telescope.  The $\gamma$-ray emission seen from
\hesssrc\ is potentially produced by a pulsar wind nebula powered by
\palfapsr; faint X-ray emission detected by \asca\ at the pulsar's
position supports this hypothesis. The inferred $\gamma$-ray
efficiency is $\epsilon_{\gamma} = L_{\gamma}/\dot{E} = 3.1$\,\%
($1-10$\,TeV, for a distance of 9\,kpc), comparable to that observed
in similar associations.
\end{abstract}


\section{INTRODUCTION}

In the last few years, approximately 40 new Galactic sources of
very-high-energy (VHE) $\gamma$-ray emission (100\,GeV$-$100\,TeV)
have been discovered using the High Energy Stereoscopic System
(HESS\footnote{See
http://www.mpi-hd.mpg.de/hfm/HESS/public/HESS\_catalog.htm for a
catalog of HESS-detected sources.}) Cherenkov telescope array
\citep[e.g.,][]{aaa+05a}.  These are an exciting new population of
sources, which give new insight into non-thermal particle acceleration
in Galactic objects such as neutron stars, supernova remnants, and
X-ray binaries.  Thus far, close to half of these sources have been
established or suggested as being associated with the pulsar wind
nebulae (PWNe) of young pulsars, either through direct detection of a
PWN or positional coincidence with a young pulsar which is presumed to
have a PWN (Table~\ref{tab:assoc}). Clearly, PWNe are now an important
and well-established Galactic source of VHE $\gamma$-rays.  Since both
young pulsars and their PWNe can be very dim, many of the Galactic
HESS sources without identified counterparts \citep[e.g.,][]{aab+08a}
may be PWNe, potentially detectable via deep radio or X-ray
observations.

In general, radio/X-ray PWNe are associated with extended HESS
sources, presumably TeV PWNe, whose peak is offset by several
arcminutes from the pulsar\footnote{Note however that this situation
does not appear to hold in the case of the youngest pulsars, e.g. the
Crab, where there is no observed offset and the TeV emission is
consistent with a point source.  This possibly demonstrates the
evolution of TeV PWNe with pulsar age.} (Table~\ref{tab:assoc}). In
some cases the offset, if any, cannot be measured because the position
of the pulsar is not known \citep[see also][]{gal07}.  These offsets
have been explained by the hypothesis that the VHE emission is from
inverse Compton scattering of ``old'' electrons, which were produced
during an earlier epoch in the pulsar's life, off of the ambient
photon field (e.g., cosmic microwave background radiation, starlight,
and infrared emission from dust; see Aharonian et al.
2005d\nocite{aab+05d}, de Jager \& Djannati-Ata\"i 2008\nocite{dd08}).
An alternative model has the $\gamma$-rays produced by the decay of
$\pi^0$ mesons, which are created by the interaction of accelerated
hadrons with nuclei in the interstellar medium \citep{has+06,hahs07}.
In both cases, asymmetric crushing of the pulsar wind by the reverse
shock of the expanding supernova remnant, in cases where the remnant
has expanded into an asymmetric density distribution in the ambient
medium, is then responsible for the offset between the pulsar and the
peak of the VHE emission \citep{dd08}.  A high pulsar proper motion
may also play a role in the offset. The VHE emission could be a key
into the earlier energetic history of these young pulsars as well as
completing a broadband (radio to TeV $\gamma$-rays) picture of the
shocked pulsar wind, whose emission would be dominated by
synchrotron-emitting electrons below $\sim 1$\,GeV and inverse Compton
emission from the scattering of these same particles above this
energy.

In this Letter, we present the discovery and subsequent timing of the
``Vela-like'' pulsar \citep[i.e. pulsars having characteristic age $\tau_c =
10-30$\,kyr and spin-down luminosity $\dot{E} \sim 10^{36}$\,ergs
s$^{-1}$; e.g.,][]{kbm+03} \palfapsr\ in the Arecibo PALFA
survey of the Galactic Plane.  This young, energetic pulsar is
positionally coincident with, and energetically capable of creating,
the VHE emission observed from the hitherto unidentified TeV source
\hesssrc\ \citep{aab+08a}.  This association suggests that some of the
other currently unidentified, extended HESS sources in the Galactic
plane may also be related to faint radio pulsars, rather than some new
source class.  An archival \asca\ image of the area around \palfapsr\
and \hesssrc\ shows a possible X-ray counterpart, cataloged as
\ascaxray\ by \citet{smk+01}, at the pulsar position; this is possibly
a synchrotron counterpart to the TeV PWN\footnote{
Furthermore, in a recent reanalysis of $EGRET$ $\gamma$-ray data,
\citet{cg08} identified the source EGR J1856+0235, which is analogous
to 3EG J1856+0114 in the 3rd $EGRET$ catalog.  \palfapsr\ is well
within the 95\% confidence region of EGR J1856+0235, suggesting the
pulsar and/or its PWN are also visible in the MeV-GeV range.  We will
investigate this further in a follow-up paper.}.

\section{OBSERVATIONS AND ANALYSIS}

\palfapsr\ was discovered in the Arecibo PALFA survey for pulsars and
radio transients (Cordes et al. 2006\nocite{cfl+06}; see also Hessels
2007\nocite{hes07}).  PALFA is using the 1.4-GHz Arecibo L-band Feed
Array (ALFA) 7-beam receiver to survey the Galactic Plane at
longitudes of $32^{\circ} < l < 77^{\circ}$ and $168^{\circ} < l <
214^{\circ}$, out to latitudes $|b| \leq 5^{\circ}$.  The relatively
high observing frequency and unequaled raw sensitivity of the Arecibo
telescope make the PALFA survey sensitive to distant, faint, and
scattered pulsars that were missed in previous surveys.  The larger
goals, design, and observational setup of the PALFA survey are
presented in detail by \citet{cfl+06}.

We found \palfapsr\ in a 268-s survey observation made on 2006 April
16. The pulsar was identified at a signal-to-noise ratio of 35 within
a few minutes of the discovery observation itself through a ``real
time'' processing pipeline, operating on data with reduced time and
spectral resolution, which is automatically run on the survey data as
it is being collected \citep{cfl+06}.  \palfapsr\ has a spin period of
81\,ms and a large dispersion measure (DM), 622\,cm$^{-3}$ pc.  We
estimate that the flux density at 1400\,MHz is $S_{1400} = 0.55 \pm
0.15$\,mJy.  We also note that \palfapsr\ shows a significant
scattering tail at 1170\,MHz ($\tau_{\rm sc} = 10 \pm 4$\,ms at this
frequency, where $\tau_{\rm sc}$ is the time constant of a one-sided
exponential fitted to the pulse profile) and would be difficult to
detect below $\sim 800$\,MHz because the scattering time scale would be
greater than the pulse period.

Immediately following the discovery of \palfapsr, we began regular
timing observations with Arecibo and the Jodrell Bank Observatory's
76-m Lovell Telescope in order to derive a precise rotational and
astrometric ephemeris.  Between the two telescopes, timing
observations were made on 148 separate days between 2006 April 24 and
2008 April 18.  At Arecibo, observations were made using both the
center pixel of the ALFA receiver at 1440\,MHz (identical setup to the
standard PALFA survey mode) and the L-Wide receiver with multiple
Wide-band Arecibo Pulsar Processor (WAPP) correlators centered at 1170,
1370, 1470, and 1570\,MHz, each with 256 lags over 100\,MHz of
bandwidth, sampled every 256\,$\mu$s. The typical integration time was
3 minutes.  The Jodrell Bank observations were made at 1402/1418\,MHz
with a $2\times 64\times 1$\,MHz incoherent filterbank system and
202\,$\mu$s sampling time. The typical integration time was 20
minutes.
The resulting phase-connected timing solution for \palfapsr\ is
presented in Table~\ref{tab:pulsar} and combines data from both
Arecibo and Jodrell Bank.  This solution shows that \palfapsr\ is a
Vela-like pulsar with a characteristic age of 21\,kyr and a spin-down
luminosity of $\dot{E} = 4.6 \times 10^{36}$\,ergs s$^{-1}$. The
solution was obtained using the {\tt TEMPO} pulsar timing
package\footnote{See http://www.atnf.csiro.au/research/pulsar/tempo.}.
The times of arrival (TOAs) can be phase-connected by fitting only for
position, DM, spin period $P$, period derivative $\dot{P}$, and period
second derivative $\ddot{P}$, though further higher-order period
derivatives are required to remove all trends in the TOAs (i.e., to
``whiten'' the residuals).  These higher-order period
derivatives, including $\ddot{P}$, are non-determinisitc and are likely
the result of timing noise, which is common in young pulsars.  Timing
noise affects the measured position of \palfapsr\ at a level exceeding
the formal {\tt TEMPO} uncertainties by an order of magnitude.
Separately fitting the TOAs from the first and second years of timing
gives positions which differ by $\sim 9^{\prime \prime}$.  We take this
as a rough measure of the true uncertainty on the timing-derived
position of \palfapsr.

The identification of \palfapsr\ as a young, energetic pulsar raises
the likelihood that it powers a radio and/or X-ray PWN \citep[for a
review, see][]{krh06}. Accordingly, we have checked source catalogs
and archival, multi-wavelength data for other sources at the radio
timing position of the pulsar.  These searches revealed that
\palfapsr\ is spatially coincident with the VHE $\gamma$-ray source
\hesssrc\ and the faint X-ray source \ascaxray.  We discuss these
possible associations in \S 3.

\section{DISCUSSION}

\begin{figure}
\plotone{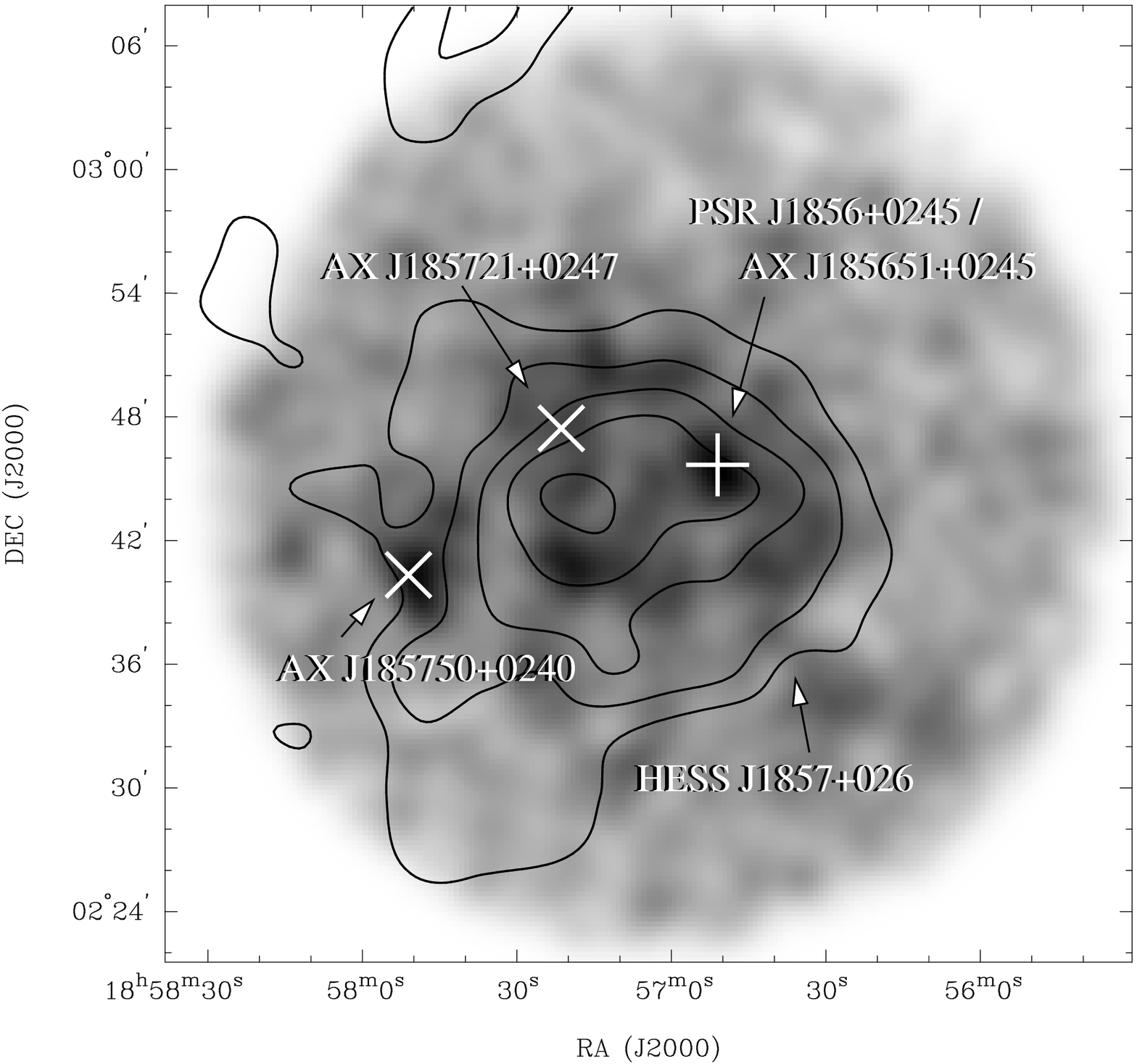} \figcaption{An \asca\ GIS image ($2-10$\,keV) of the
region surrounding \palfapsr.  This image has not been background
subtracted or corrected for vignetting.  The greyscale shows the GIS
image smoothed with a $2^{\prime}$ (FWHM) Gaussian and scaled to bring
out possible faint extended structure.  The contours show significance
levels of \hesssrc\ from $5-9$\,$\sigma$, in steps of 1\,$\sigma$
\citep[from][]{aab+08a}.  The plus marks \palfapsr\ and \ascaxray\ and
is much larger than the uncertainty on the pulsar's position.  Two
other nearby sources detected by \citet{smk+01}, AX J185721+0247 and
AX J185750+0240, are also marked.  It is possible that AX J185721+0247
shows no excess in this image because it is faint and soft. 
\label{fig:xray}}
\end{figure}

\palfapsr\ is spatially coincident with the unidentified VHE
$\gamma$-ray source \hesssrc\ discovered by \citet[][see
Fig.~\ref{fig:xray}]{aab+08a}.  Here we show that \palfapsr\ is
energetically capable of powering \hesssrc\ and that this association
has similar characteristics to the other pulsar/VHE associations in
the literature (Table~\ref{tab:assoc}).

The estimated distance to \palfapsr, based on its DM and the NE2001
electron density model of the Galaxy \citep{cl02}, is $\sim 9$\,kpc.
The uncertainty on this distance is not well-defined, but in some
cases the model can be off by as much as a factor of $\sim 2-3$.
Adopting the DM-distance of 9\,kpc gives a large spin-down flux of
$\dot{E}/d^2 = 5.7/d^2_9 \times 10^{34}$\,ergs s$^{-1}$ kpc$^{-2}$,
where $d_9$ is the true distance scaled to 9\,kpc. \citet{chh+07} find
that statistically 70\,\% of pulsars with $\dot{E}/d^2 \gtrsim
10^{35}$\,ergs s$^{-1}$ kpc$^{-2}$ are visible as VHE $\gamma$-ray
sources. \palfapsr\ is very close to this limit and may exceed it if
its distance is overestimated.  Hence, based on its energetics alone,
it is likely to be visible as a VHE $\gamma$-ray source.
\hesssrc\ has a photon index $\Gamma = 2.2$ and $1-10$\,TeV flux
$F_{\rm VHE} = 1.5 \times 10^{-11}$\,ergs cm$^{-2}$ s$^{-1}$ (about
15\,\% of the Crab's flux in this energy range); these spectral
parameters are similar to those measured for the other HESS sources
identified with PWNe.  Given the spin-down luminosity
of \palfapsr, this suggests an efficiency $\epsilon_{\gamma} =
L_{\gamma}/\dot{E} = 3.1d^2_9$\,\% ($1-10$\,TeV), comparable to what
is seen in other proposed associations (Table~\ref{tab:assoc}).

\palfapsr\ is offset from the centroid of \hesssrc, $\alpha =
18^{\rm{h}}57^{\rm{m}}11^{\rm{s}}$, $\delta =
+02^{\circ}40^{\prime}00^{\prime \prime}$ (J2000, there is a
$3^{\prime}$ statistical uncertainty on this position), by
$8^{\prime}$ (Fig.~\ref{fig:xray}).  As discussed in \S 1, this is
most likely explained by asymmetric confinement of the pulsar wind.
This interpretation is supported by \palfapsr's offset location on the
side of \hesssrc\ that appears somewhat compressed (i.e., there is a
steep gradient in the $\gamma$-rays) compared with the rest of the
nebula.  If, however, the offset of the VHE emission is due primarily
to the proper motion of \palfapsr, then the direction and rough
magnitude of this motion are predictable.  Based on its characteristic
age and offset from the centroid of \hesssrc, \palfapsr's proper
motion should be roughly 23\,mas yr$^{-1}$ (transverse velocity
$v_{\rm t} = 970d_9$\,km s$^{-1}$), to the north-west, assuming that
the centroid of \hesssrc\ marks the birthplace of the pulsar.  The
velocity is very large, but not unprecedented for a pulsar
\citep{cvb+05}.  Of course, the velocity will be smaller if the
distance to the source is overestimated, or if the offset is at least
partially due to an asymmetrically confined pulsar wind. Detecting
this proper motion via timing or interferometry would elucidate this
further and may be possible in the coming years, though timing noise
and the low flux of the pulsar will make this difficult. 

\palfapsr\ and \hesssrc\ are also coincident with the faint \asca\
X-ray source \ascaxray\ reported by \citet[][see
Fig.~\ref{fig:xray}]{smk+01}. \ascaxray\ was found $3^{\prime}$ off
axis in observations from 1998 April 6 (sequence number 56003000).  It
was detected only in the hard band ($2-10$\,keV) of the Gas Imaging
Spectrometers (GISs), with a count rate of 2.6\,ks$^{-1}$ GIS$^{-1}$
and a significance of 4.3\,$\sigma$.  The exposure was $\sim 13$\,ks
for each of GIS2 and GIS3.  \palfapsr\ and \ascaxray\ (J2000 position:
$\alpha = 18^{\rm{h}}56^{\rm{m}}50^{\rm{s}}$, $\delta =
+02^{\circ}46^{\prime}$) are spatially coincident to within the
$1^{\prime}$ positional uncertainty of sources in the \citet{smk+01}
catalog.  Although the signal-to-noise ratio of the detection of
\ascaxray\ is modest, its exact spatial coincidence with a young
pulsar of relatively high spin-down flux argues that this source is
real.
Most of the previously established associations of HESS sources with
young pulsars also have known X-ray synchrotron PWNe.  \ascaxray\
could be an X-ray PWN created by \palfapsr.  Using
CXCPIMMS\footnote{Available at
http://cxc.harvard.edu/toolkit/pimms.jsp.}, with an assumed absorbed
power law spectrum with column density $N_{\rm H} = 1 \times
10^{22}$\,cm$^{-2}$ (roughly the total Galactic contribution in this
direction) and photon index $\Gamma = 2$ (typical for X-ray PWNe), we
find that \ascaxray\ has an unabsorbed flux ($2-10$\,keV) of $1.6
\times 10^{-13}$\,ergs s$^{-1}$ cm$^{-2}$.  This corresponds to an
efficiency for the conversion of spin-down energy into X-rays
$\epsilon_{\rm X} = L_{\rm X}/\dot{E} = 0.03d^2_9$\,\% ($2-10$\,keV)
that falls into the observed range for Vela-like pulsars
\citep{pccm02}.  There are two additional nearby sources detected by
\citet{smk+01}: AX J185721+0247 and AXJ185750+0240.  It is possible
that these are part of some extended X-ray emission related with
\hesssrc, though deeper X-ray observations are needed to investigate
this (see below).  

There are four short \swift\ observations of the region containing
\palfapsr, including two observations specifically targeting
\ascaxray. The deepest of these, from 2007 March 13 (observation ID
36183002), is a 4.1-ks on-axis exposure with the \swift\ X-ray
Telescope (XRT). This observation does not show any significant
emission at the pulsar position; within a circular extraction region
of radius $1^{\prime}$ around the pulsar, the background subtracted
number of counts is $1.1^{+4.0}_{-2.8}$.  Using the aforementioned
count rate of \ascaxray, along with the same assumed spectrum, the
predicted $0.2-10$\,keV count rate for XRT from CXCPIMMS is $\sim
2$\,counts/ks. Thus, in 4.1\,ks, there should have been $\sim
8$\,counts, consistent with the 2-$\sigma$ upper limit derived from
the data.  Thus, if \ascaxray\ is predominantly a point source, it was
only at the limit of detectability in this observation.  If it is
predominantly an extended nebula, then it would not have been
detectable in such a short exposure. Clearly, future, deeper
observations, like those that recently discovered a likely X-ray PWN
associated with HESS J1718$-$385/PSR J1718$-$3825 \citep{hfc+07}, will
be needed to determine the nature of this candidate X-ray PWN.  We
have been granted a $\sim 60$-ks {\it XMM-Newton} observation of
\palfapsr\ and will present an analysis of those data in a follow-up
paper.  

At least half of the known HESS/pulsar associations are accompanied by
radio emission classified as a PWN, a notable exception being HESS
J1825$-$137/PSR B1823$-$13. Only a few of the extended HESS sources,
e.g. HESS J0835$-$455, HESS J1813$-$178, and HESS J1640$-$465, are
known to be accompanied by a supernova remnant (SNR). We have checked
available radio imaging data for signs of a PWN or SNR. There is some
faint, extended emission in the vicinity of \palfapsr\ visible in
1.4-GHz VLA Galactic Plane Survey data \citep[VGPS,][]{std+06}, though
nothing that is clearly indicative of a PWN or SNR.  The surface
brightness limit from VGPS is $\sim 1.8 \times 10^{5}$\,Jy sr$^{-1}$
at 1.4\,GHz.  It is certainly not uncommon for Vela-like pulsars to
have faint or no known radio nebula \citep[e.g. PSR
B1823$-$13,][]{bgl89,gsk+03}.  Roughly a third of the cataloged
Galactic SNRs \citep{gre04} are fainter than the surface brightness
limit achieved by the VGPS. Deep, dedicated radio imaging observations
of \palfapsr\ are necessary to investigate this further. Higher
resolution 1.4-GHz data from the MAGPIS survey (Helfand et al.
2006\nocite{hbw+06}; in the vicinity of \palfapsr, this survey has a
sensitivity of 0.2\,mJy/beam at an angular resolution of $6^{\prime
\prime}$) reveal no point source which can be definitively associated
with \palfapsr, as expected given the flux density and positional
uncertainty of the pulsar.

\acknowledgements The Arecibo observatory, a facility of the NAIC, is
operated by Cornell University in a cooperative agreement with the
National Science Foundation.  We thank Karl Kosack and the HESS
collaboration for providing the $\gamma$-ray image of \hesssrc. This
work was supported by NSERC (CGS-D, PDF, and Discovery grants), the
Canadian Space Agency, the Australian Research Council, FQRNT, the
Canadian Institute for Advanced Research, the Canada Research Chairs
Program, the McGill University Lorne Trottier Chair in Astrophysics
and Cosmology, and NSF grants AST-0647820 and AST-0545837.


\bibliographystyle{apj}

\begin{thebibliography}{43}
\expandafter\ifx\csname natexlab\endcsname\relax\def\natexlab#1{#1}\fi

\bibitem[{{Aharonian} {et~al.}(2005{\natexlab{a}}){Aharonian}, {Akhperjanian},
  {Aye}, {Bazer-Bachi}, {Beilicke}, {Benbow}, {Berge}, {Berghaus}, {Bernlohr},
  {Boisson}, {Bolz}, {Borgmeier}, {Braun}, \& {Breitling}}]{aaa+05a}
  {Aharonian}, F.~A., et al. (HESS Collaboration) 2005{\natexlab{a}}, Science, 307, 1938

\bibitem[{{Aharonian} {et~al.}(2005{\natexlab{b}}){Aharonian}, {Akhperjanian},
  {Aye}, {Bazer-Bachi}, {Beilicke}, {Benbow}, {Berge}, {Berghaus},
  {Bernl{\"o}hr}, {Boisson}, {Bolz}, {Borgmeier}, {Braun}, {Breitling},
  {Brown}, {Bussons Gordo}, \& {Chadwick}}]{aaa+05e}
---. 2005{\natexlab{b}}, \aap, 432, L25

\bibitem[{{Aharonian} {et~al.}(2005{\natexlab{c}}){Aharonian}, {Akhperjanian},
  {Aye}, {Bazer-Bachi}, {Beilicke}, {Benbow}, {Berge}, {Berghaus}, {Bernlohr},
  {Boisson}, {Bolz}, {Braun}, {Breitling}, \& {Brown}}]{aaa+05c}
---. 2005{\natexlab{c}}, \aap, 435, L17

\bibitem[{{Aharonian} {et~al.}(2005{\natexlab{d}}){Aharonian}, {Akhperjanian},
  {Bazer-Bachi}, {Beilicke}, {Benbow}, {Berge}, {Bernlohr}, {Boisson}, {Bolz},
  {Borrel}, \& {Braun}}]{aab+05d}
---. 2005{\natexlab{d}}, \aap, 442, L25

\bibitem[{{Aharonian} {et~al.}(2006{\natexlab{a}}){Aharonian}, {Akhperjanian},
  {Bazer-Bachi}, {Beilicke}, {Benbow}, {Berge}, {Bernlohr}, {Boisson}, {Bolz},
  {Borrel}, \& {Braun}}]{aab+06b}
---. 2006{\natexlab{a}}, \aap, 456, 245

\bibitem[{{Aharonian} {et~al.}(2006{\natexlab{b}}){Aharonian}, {Akhperjanian},
  {Bazer-Bachi}, {Beilicke}, {Benbow}, {Berge}, {Bernlohr}, {Boisson}, {Bolz},
  {Borrel}, \& {Braun}}]{aab+06a}
---. 2006{\natexlab{b}}, \aap, 448, L43

\bibitem[{{Aharonian} {et~al.}(2006{\natexlab{c}}){Aharonian}, {Akhperjanian},
  {Bazer-Bachi}, {Beilicke}, {Benbow}, {Berge}, {Bernlohr}, {Boisson}, {Bolz},
  {Borrel}, \& {Braun}}]{aab+06c}
---. 2006{\natexlab{c}}, \aap, 457, 899

\bibitem[{{Aharonian} {et~al.}(2006{\natexlab{d}}){Aharonian}, {Akhperjanian},
  {Bazer-Bachi}, {Beilicke}, {Benbow}, {Berge}, {Bernl{\"o}hr}, {Boisson},
  {Bolz}, {Borrel}, {Braun}, {Breitling}, {Brown}, \& {Chadwick}}]{aab+06e}
---. 2006{\natexlab{d}}, \apj, 636, 777

\bibitem[{{Aharonian} {et~al.}(2007{\natexlab{a}}){Aharonian}, {Akhperjanian},
  {Bazer-Bachi}, {Behera}, {Beilicke}, {Benbow}, {Berge}, {Bernlohr},
  {Boisson}, {Bolz}, {Borrel}, {Braun}, {Brion}, \& {Brown}}]{aab+07b}
---. 2007{\natexlab{a}}, \aap, 472, 489

\bibitem[{{Aharonian} {et~al.}(2007{\natexlab{b}}){Aharonian}, {Akhperjanian},
  {Bazer-Bachi}, {Beilicke}, {Benbow}, {Berge}, {Bernl{\"o}hr}, {Boisson},
  {Bolz}, {Borrel}, {Braun}, {Brown}, {B{\"u}hler}, {B{\"u}sching}, {Carrigan},
  {Chadwick}, {Chounet}, {Coignet}, {Cornils}, {Costamante}, \&
  {Degrange}}]{aab+07c}
---. 2007{\natexlab{b}}, \apj, 661, 236

\bibitem[{{Aharonian} {et~al.}(2008{\natexlab{a}}){Aharonian}, {Akhperjanian},
  {Barres de Almeida}, {Bazer-Bachi}, {Behera}, {Beilicke}, {Benbow},
  {Bernlohr}, {Boisson}, {Bolz}, {Borrel}, \& {Braun}}]{aab+08b}
---. 2008{\natexlab{a}}, preprint (astro-ph/0802.3841)

\bibitem[{{Aharonian} {et~al.}(2008{\natexlab{b}}){Aharonian}, {Akhperjanian},
  {Barres de Almeida}, {Bazer-Bachi}, {Behera}, {Beilicke}, {Benbow},
  {Bernlohr}, {Boisson}, {Bolz}, {Borrel}, \& {Braun}}]{aab+08a}
---. 2008{\natexlab{b}}, \aap, 477, 353

\bibitem[{{Braun} {et~al.}(1989){Braun}, {Goss}, \& {Lyne}}]{bgl89}
{Braun}, R., {Goss}, W.~M., \& {Lyne}, A.~G. 1989, \apj, 340, 355

\bibitem[{{Brogan} {et~al.}(2005){Brogan}, {Gaensler}, {Gelfand}, {Lazendic},
  {Lazio}, {Kassim}, \& {McClure-Griffiths}}]{bgg+05}
{Brogan}, C.~L., {Gaensler}, B.~M., {Gelfand}, J.~D., {Lazendic}, J.~S.,
  {Lazio}, T.~J.~W., {Kassim}, N.~E., \& {McClure-Griffiths}, N.~M. 2005,
  \apjl, 629, L105

\bibitem[{{Carrigan} {et~al.}(2007){Carrigan}, {Hinton}, {Hofmann}, {Kosack},
  {Lohse}, {Reimer}, \& {the HESS~Collaboration}}]{chh+07}
{Carrigan}, S., {Hinton}, J.~A., {Hofmann}, W., {Kosack}, K., {Lohse}, T.,
  {Reimer}, O., \& {the HESS~Collaboration}. 2007, preprint
  (astro-ph/0709.4094)

\bibitem[{{Casandjian} \& {Grenier}(2008)}]{cg08}
  {Casandjian}, J.-M. \& {Grenier}, I.~A. 2008, preprint
  (astro-ph/0806.0113)

\bibitem[{{Chatterjee} {et~al.}(2005){Chatterjee}, {Vlemmings}, {Brisken},
  {Lazio}, {Cordes}, {Goss}, {Thorsett}, {Fomalont}, {Lyne}, \&
  {Kramer}}]{cvb+05}
  {Chatterjee}, S., et al. 2005, \apjl, 630, L61

\bibitem[{{Cordes} {et~al.}(2006){Cordes}, {Freire}, {Lorimer}, {Camilo},
  {Champion}, {Nice}, {Ramachandran}, {Hessels}, {Vlemmings}, {van Leeuwen},
  {Ransom}, {Bhat}, {Arzoumanian}, {McLaughlin}, {Kaspi}, {Kasian}, {Deneva},
  {Reid}, {Chatterjee}, {Han}, {Backer}, {Stairs}, {Deshpande}, \&
  {Faucher-Gigu{\`e}re}}]{cfl+06}
  {Cordes}, J.~M., et al. 2006, \apj, 637, 446

\bibitem[{{Cordes} \& {Lazio}(2002)}]{cl02}
  {Cordes}, J.~M. \& {Lazio}, T.~J.~W. 2002, preprint (astro-ph/0207156)

\bibitem[{{de Jager} \& {Djannati-Ata$\ddot{\rm \i}$}(2008)}]{dd08}
  {de Jager}, O.~C. \& {Djannati-Ata$\ddot{\rm \i}$}, A. 2008, preprint
  (astro-ph/0803.0116)

\bibitem[{{Djannati-Ata$\ddot{\i}$} {et~al.}(2007){Djannati-Ata$\ddot{\i}$},
  {De Jager}, {Terrier}, {Gallant}, {Hoppe}, \& {the
  HESS~Collaboration}}]{ddt+07}
  {Djannati-Ata$\ddot{\i}$}, A., {De Jager}, O.~C., {Terrier}, R., {Gallant},
  Y.~A., {Hoppe}, S., \& {the HESS~Collaboration}. 2007, preprint
  (astro-ph/0710.2247)

\bibitem[{{Funk} {et~al.}(2007{\natexlab{a}}){Funk}, {Hinton}, {Moriguchi},
  {Aharonian}, {Fukui}, {Hofmann}, {Horns}, {P$\ddot{\rm u}$hlhofer}, {Reimer},
  {Rowell}, {Terrier}, {Vink}, \& {Wagner}}]{fhm+07}
  {Funk}, S., et al. 2007{\natexlab{a}}, \aap, 470, 249

\bibitem[{{Funk} {et~al.}(2007{\natexlab{b}}){Funk}, {Hinton}, {P$\ddot{\rm
  u}$hlhofer}, {Aharonian}, {Hofmann}, {Reimer}, \& {Wagner}}]{fhp+07}
  {Funk}, S., {Hinton}, J.~A., {P$\ddot{\rm u}$hlhofer}, G., {Aharonian}, F.~A.,
  {Hofmann}, W., {Reimer}, O., \& {Wagner}, S. 2007{\natexlab{b}}, \apj, 662,
  517

\bibitem[{{Gaensler} {et~al.}(2003){Gaensler}, {Schulz}, {Kaspi}, {Pivovaroff},
  \& {Becker}}]{gsk+03}
  {Gaensler}, B.~M., {Schulz}, N.~S., {Kaspi}, V.~M., {Pivovaroff}, M.~J., \&
  {Becker}, W.~E. 2003, \apj, 588, 441

\bibitem[{{Gallant}(2007)}]{gal07}
{Gallant}, Y.~A. 2007, \apss, 309, 197

\bibitem[{{Gotthelf} \& {Halpern}(2008)}]{gh08}
{Gotthelf}, E.~V. \& {Halpern}, J.~P. 2008, preprint (astro-ph/0803.1361)

\bibitem[{{Green}(2004)}]{gre04}
{Green}, D.~A. 2004, Bulletin of the Astronomical Society of India, 32, 335

\bibitem[{{Helfand} {et~al.}(2006){Helfand}, {Becker}, {White}, {Fallon}, \&
  {Tuttle}}]{hbw+06}
{Helfand}, D.~J., {Becker}, R.~H., {White}, R.~L., {Fallon}, A., \& {Tuttle},
  S. 2006, \aj, 131, 2525

\bibitem[{{Helfand} {et~al.}(2007){Helfand}, {Gotthelf}, {Halpern}, {Camilo},
  {Semler}, {Becker}, \& {White}}]{hgh+07}
{Helfand}, D.~J., {Gotthelf}, E.~V., {Halpern}, J.~P., {Camilo}, F., {Semler},
  D.~R., {Becker}, R.~H., \& {White}, R.~L. 2007, \apj, 665, 1297

\bibitem[{{Hessels}(2007)}]{hes07}
{Hessels}, J.~W.~T. 2007, PhD thesis, McGill University

\bibitem[{{Hinton} {et~al.}(2007){Hinton}, {Funk}, {Carrigan}, {Gallant}, {de
  Jager}, {Kosack}, {Lemi{\`e}re}, \& {P{\''u}hlhofer}}]{hfc+07}
  {Hinton}, J.~A., et al. 2007, \aap, 476, L25

\bibitem[{{Horns} {et~al.}(2007){Horns}, {Aharonian}, {Hoffmann}, \&
  {Santangelo}}]{hahs07}
{Horns}, D., {Aharonian}, F., {Hoffmann}, A.~I.~D., \& {Santangelo}, A. 2007,
  \apss, 309, 189

\bibitem[{{Horns} {et~al.}(2006){Horns}, {Aharonian}, {Santangelo}, {Hoffmann},
  \& {Masterson}}]{has+06}
{Horns}, D., {Aharonian}, F., {Santangelo}, A., {Hoffmann}, A.~I.~D., \&
  {Masterson}, C. 2006, \aap, 451, L51

\bibitem[{{Kargaltsev} \& {Pavlov}(2007)}]{kp07}
{Kargaltsev}, O. \& {Pavlov}, G.~G. 2007, \apj, 670, 655

\bibitem[{{Kargaltsev} {et~al.}(2007){Kargaltsev}, {Pavlov}, \&
  {Garmire}}]{kpg07}
{Kargaltsev}, O., {Pavlov}, G.~G., \& {Garmire}, G.~P. 2007, \apj, 670, 643

\bibitem[{{Kaspi} {et~al.}(2006){Kaspi}, {Roberts}, \& {Harding}}]{krh06}
{Kaspi}, V.~M., {Roberts}, M.~S.~E., \& {Harding}, A.~K. 2006, Compact stellar
  X-ray sources, 279

\bibitem[{{Kramer} {et~al.}(2003){Kramer}, {Bell}, {Manchester}, {Lyne},
  {Camilo}, {Stairs}, {D'Amico}, {Kaspi}, {Hobbs}, {Morris}, {Crawford},
  {Possenti}, {Joshi}, {McLaughlin}, {Lorimer}, \& {Faulkner}}]{kbm+03}
  {Kramer}, M., et al. 2003, \mnras, 342, 1299

\bibitem[{{Landi} {et~al.}(2007){Landi}, {de Rosa}, {Dean}, {Bassani},
  {Ubertini}, \& {Bird}}]{ldd+07}
  {Landi}, R., {de Rosa}, A., {Dean}, A.~J., {Bassani}, L., {Ubertini}, P., \&
  {Bird}, A.~J. 2007, \mnras, 380, 926

\bibitem[{{Mukherjee} \& {Halpern}(2005)}]{mh05}
{Mukherjee}, R. \& {Halpern}, J.~P. 2005, \apj, 629, 1017

\bibitem[{{Ng} {et~al.}(2005){Ng}, {Roberts}, \& {Romani}}]{nrr05}
{Ng}, C.-Y., {Roberts}, M.~S.~E., \& {Romani}, R.~W. 2005, \apj, 627, 904

\bibitem[{{Possenti} {et~al.}(2002){Possenti}, {Cerutti}, {Colpi}, \&
  {Mereghetti}}]{pccm02}
{Possenti}, A., {Cerutti}, R., {Colpi}, M., \& {Mereghetti}, S. 2002, \aap,
  387, 993

\bibitem[{{Stil} {et~al.}(2006){Stil}, {Taylor}, {Dickey}, {Kavars}, {Martin},
  {Rothwell}, {Boothroyd}, {Lockman}, \& {McClure-Griffiths}}]{std+06}
  {Stil}, J.~M., et al. 2006, \aj, 132, 1158

\bibitem[{{Sugizaki} {et~al.}(2001){Sugizaki}, {Mitsuda}, {Kaneda},
  {Matsuzaki}, {Yamauchi}, \& {Koyama}}]{smk+01}
{Sugizaki}, M., {Mitsuda}, K., {Kaneda}, H., {Matsuzaki}, K., {Yamauchi}, S.,
  \& {Koyama}, K. 2001, \apjs, 134, 77

\end{thebibliography}


\clearpage

\begin{deluxetable}{lccccccccc}
\tablewidth{0pt}
\tabletypesize{\scriptsize}
\tablecaption{HESS VHE $\gamma$-ray Sources Possibly Associated with PWNe \label{tab:assoc}}
\tablehead{
\colhead{HESS} & 
\colhead{Size\tablenotemark{a}} & 
\colhead{Pulsar /} & 
\colhead{Offset} & 
\colhead{$P_{\rm spin}$} & 
\colhead{$\tau_{\rm c}$} & 
\colhead{$\dot{E}$} & 
\colhead{$d$\tablenotemark{b}} & 
\colhead{$L_{\gamma}/\dot{E}$} &
\colhead{Assoc.} \\
\colhead{Source} & 
\colhead{(arcmin)} & 
\colhead{PWN} & 
\colhead{(arcmin)} & 
\colhead{(ms)} & 
\colhead{(kyr)} & 
\colhead{$\times 10^{36}$} &
\colhead{(kpc)} & 
\colhead{($1-10$\,TeV)} &
\colhead{Ref.} \\
&
&
&
&
&
&
\colhead{(ergs s$^{-1}$)} & 
&
\colhead{(\%)} &
}
\startdata
J0534+220 & {\ldots} & B0531+21 & {\ldots} & 33 & 1 & 460 & 2 & 0.009 & 1 \\
J0835$-$455 & 26 & B0833$-$45 & 18 & 89 & 11 & 6.9 & 0.3 & 0.008 & 2 \\
J0852$-$463 & 60 & J0855$-$4644 & 43 & 65 & 141 & 1.1 & {\it 2} & 3 & 3 \\
J1303$-$631 & 10 & J1301$-$6305 & 11 & 185 & 11 & 1.7 & {\it 7} & 5 & 4 \\
J1418$-$609\tablenotemark{c} & 4 & G313.3+0.1 & 8 & {\ldots} & {\ldots} & {\ldots} & {\ldots} & {\ldots} & 5 \\
J1420$-$607 & 4 & J1420$-$6048 & 3 & 68 & 13 & 10 & {\it 6} & 0.4 & 5 \\
J1514$-$591 & 6 & B1509$-$58 & 2 & 151 & 16 & 18 & {\it 4} & 0.2 & 6 \\
J1616$-$508 & 8 & J1617$-$5055 & 10 & 69 & 8 & 16 & {\it 7} & 0.6 & 7 \\
J1640$-$465 & 3 & G338.3$-$0.0 & {\ldots} & {\ldots}  & {\ldots} & {\ldots} & {\ldots} & {\ldots} & 8 \\
J1702$-$420 & 18 & J1702$-$4128 & 35 & 182 & 55 & 0.34 & {\it 5} & 5 & 9 \\
J1718$-$385 & 9 & J1718$-$3825 & 8 & 75 & 90 & 1.3 & {\it 4} & 0.4 & 10 \\
J1747$-$281 & {\ldots} & G0.9+0.1 & {\ldots} & {\ldots} & {\ldots} & {\ldots} & {\ldots} & {\ldots} & 11 \\
J1804$-$216 & 12 & B1800$-$21 & 10 & 134 & 16 & 2.2 & {\it 4} & 2 & 12 \\
J1809$-$193 & 32 & J1809$-$1917 & 12 & 83 & 51 & 1.8 & {\it 4} & 1 & 13 \\
J1813$-$178 & 2 & G12.8$-$0.0 & {\ldots} & {\ldots} & {\ldots} & {\ldots} & {\ldots} & {\ldots} & 14 \\
J1825$-$137 & 10 & B1823$-$13 & 11 & 101 & 21 & 2.8 & {\it 4} & 1 & 15 \\
J1833$-$105 & {\ldots} & J1833$-$1034 & {\ldots} & 62 & 5 & 34 & {\it 3} & 0.008 & 16 \\
J1837$-$069 & 7 & J1838$-$0655 & 6 & 70 & 23 & 5.5 & 7 & 1 & 17 \\
J1846$-$029 & {\ldots} & J1846$-$0258 & {\ldots} & 326 & 1 & 8.1 & 6 & 0.1 & 16 \\
J1857+026 & 7 & J1856+0245 & 8 & 81 & 21 & 4.6 & {\it 9} & 3 & 18 \\    
J1912+101 & 16 & J1913+1011 & 9 & 36 & 170 & 2.9 & {\it 5} & 0.6 & 19 \\
\enddata
\tablecomments{These are possible associations found in the
literature.  They are not all equally well-established.  References: 1: \citet{aab+06c}, 2: \citet{aab+06a}, 3:
\citet{aab+07c}, 4: there are difficulties with the association of this HESS source with PSR
J1301$-$6305, see \citet{mh05}, 5: \citet{aab+06b}, 6: \citet{aaa+05c},
7: \citet{ldd+07}, 8: \citet{fhp+07}, 9: \citet{aab+06e}, 10:
\citet{aab+07b}, 11: \citet{aaa+05e}, 12: \citet{aab+06e},
though the situation is unclear, see \citet{kpg07}, 13: \citet{kp07},
14: \citet{bgg+05,hgh+07,fhm+07}, 15: \citet{aab+05d}, 16:
\citet{ddt+07}, 17: \citet{gh08}, 18: this paper, 19: \citet{aab+08b}}
\tablenotetext{a}{This is the approximate source radius, taken from
http://www.mpi-hd.mpg.de/hfm/HESS/public/HESS\_catalog.htm when
available.  Some sources are consistent with being point-like and thus
no size is given.}
\tablenotetext{b}{Distances in italics are obtained from the DM and the
\citet{cl02} model.}
\tablenotetext{c}{No known radio pulsar, though possible X-ray pulsations with a period of
108\,ms were reported by \citet{nrr05}.}
\end{deluxetable}

\clearpage

\begin{deluxetable}{l c}
\tabletypesize{\scriptsize}
\tablewidth{0pt}
\tablecaption{Measured and Derived Parameters for \palfapsr \label{tab:pulsar}}
\tablehead{\colhead{Parameter} & \colhead{Value}}

\startdata
\multicolumn{2}{c}{Observation and Data Reduction Parameters}\\
\hline
Period epoch (MJD) \dotfill                & 54167 \\
Start time (MJD, Arecibo/Jodrell) \dotfill & 53849/53877 \\
End time (MJD, Arecibo/Jodrell) \dotfill   & 54570/54541 \\
Number of TOAs (Arecibo/Jodrell) \dotfill  & 78/134 \\
TOA rms (ms) \dotfill                      & 1.8 \\
\hline
\multicolumn{2}{c}{Timing Parameters}\\
\hline

Right ascension\tablenotemark{a} $\alpha$ (J2000) \dotfill &
$18^{\rm h}56^{\rm m}50\fs 80\pm 0\fs 02 \pm 1\fs 20$ \\
Declination\tablenotemark{a} $\delta$ (J2000) \dotfill &
$+02\arcdeg 45\arcmin 52\farcs 2\pm 0\farcs 4 \pm 8\farcs 0$ \\
Galactic longitude $l$ \dotfill & 36\fdg 008 \\
Galactic latitude $b$ \dotfill & +0\fdg 058 \\
Pulse period $P$ (s) \dotfill & 0.080902591336(5) \\
Period derivative $\dot P$ \dotfill & $6.2179(2) \times 10^{-14}$ \\                                   
Period second derivative\tablenotemark{b} $\ddot P$ (s$^{-1}$) \dotfill & $ -2.2(1) \times 10^{-24}$ \\
Dispersion measure DM\tablenotemark{c} (cm$^{-3}$ pc) \dots & 622(2) \\
\hline
\multicolumn{2}{c}{Derived Parameters}\\
\hline
Distance $d$ (kpc) \dotfill & $\sim 9$ \\
Spin-down luminosity $\dot{E}$ (ergs s$^{-1}$) \dotfill & $4.6 \times 10^{36}$ \\
Surface dipole magnetic field $B$ (G) \dotfill & $2.3 \times 10^{12}$ \\
Characteristic age $\tau_{c}$ (kyr) \dotfill & 21 \\

\enddata         

\tablecomments{Unless otherwise indicated, figures in parentheses are
the 1-$\sigma$ uncertainty on the least-significant digits quoted,
from {\tt TEMPO}.}

\tablenotetext{a}{Includes the statistical error
from {\tt TEMPO} plus the much larger systematic error due to timing
noise.}

\tablenotetext{b}{Value is contaminated by timing noise.}

\tablenotetext{c}{Uncertainty includes contaminating effects due to scattering.}

\end{deluxetable}


\clearpage

\end{document}